\documentclass[
reprint,
aps,
superscriptaddress,
nofootinbib,
nobibnotes,
twocolumn]{revtex4-2}

\usepackage{xcolor}
\usepackage{graphicx}
\graphicspath{{./fig/}}
\usepackage{amsmath,amssymb,amsfonts,mathrsfs,bm}
\usepackage{slashed}
\usepackage[%
colorlinks=true,
linkcolor=blue,
citecolor=blue,
]{hyperref}
\usepackage{braket}

\begin{document}
	\title{The curvature perturbations and induced gravitational waves induced by the first-order phase transition during reheating}
	
	\author{Xiao-Bin Sui}
	\email{suixiaobin21@mails.ucas.ac.cn}
	\affiliation{School of Fundamental Physics and Mathematical Sciences, Hangzhou Institute for Advanced Study, University of Chinese Academy of Sciences (HIAS-UCAS), Hangzhou 310024, China}
	\affiliation{CAS Key Laboratory of Theoretical Physics, Institute of Theoretical Physics, Chinese Academy of Sciences, Beijing 100190, China}
	\affiliation{University of Chinese Academy of Sciences, Beijing 100049, China}
	
	\author{Jing Liu}
	\email{liujing@ucas.ac.cn}
	\affiliation{International Centre for Theoretical Physics Asia-Pacific, University of Chinese Academy of Sciences, Beijing 100190, China}
	\affiliation{Taiji Laboratory for Gravitational Wave Universe (Beijing/Hangzhou), University of Chinese Academy of Sciences, Beijing 100049, China}
	
	
	\author{Rong-Gen Cai}
	\email{cairg@itp.ac.cn}
	\affiliation{Institute of Fundamental Physics and Quantum Technology, Ningbo University, Ningbo, 315211, China}
	\affiliation{CAS Key Laboratory of Theoretical Physics, Institute of Theoretical Physics, Chinese Academy of Sciences, Beijing 100190, China}
	\affiliation{School of Fundamental Physics and Mathematical Sciences, Hangzhou Institute for Advanced Study, University of Chinese Academy of Sciences (HIAS-UCAS), Hangzhou 310024, China}
	
	\begin{abstract}
		
		We propose a novel mechanism where a first-order phase transition modulates the decay rate of a massive field. This modulation, even if the scalar field has negligible energy density, subsequently generates an observable stochastic gravitational-wave background. The stochastic nature of bubble nucleation leads to the asynchrony of phase transitions, generating superhorizon-scale density perturbations through spatial variations in the decay rate $\Gamma$. These perturbations subsequently source second-order gravitational waves with peak amplitudes governed by the phase transition parameter \(\beta/H_*\) and decay rate $\Gamma$. 
		We apply this mechanism in the reheating scanario where the decay rate of inflaton are modulated by the scalar field that undergoes a first-order phase transition.
		Numerical calculations reveal that the gravitational wave energy spectrum typically reaches \(\Omega_{\text{GW}} \sim 10^{-10}\), demonstrating prospects for detection by space-based interferometers like LISA, TianQin and Taiji. 
		This work establishes a new approach to probe phase transition processes in the early Universe without requiring significant vacuum energy release.

	\end{abstract}

	\maketitle
	
	\section{Introduction}
	Gravitational waves (GWs), ripples in spacetime predicted by Einstein's general theory of relativity, provide a powerful new window into the Universe~\cite{LIGOScientific:2016aoc,Bian:2021ini,Cai:2017cbj}.
	In addition to those generated by the violent dynamics of astrophysical objects, the stochastic GW backgrounds produced in the early Universe have attracted significant attention~\cite{Ananda:2006af,Cai:2018dig,Cai:2019amo,Cai:2019cdl,Domenech:2021ztg,Kohri:2018awv,Saito:2008jc,Jiang:2024aju,Li:2024lxx}, and these backgrounds hold vital information about the evolution of early Universe~\cite{Lozanov:2019ylm,Liu:2023tmv,Bhaumik:2022pil,Sui:2024nip,Sui:2024grm,KAGRA:2021kbb,Zhou:2024yke,Inomata:2018epa}. 
	One of the violent processes in the early Universe that is expected to produce observable relics of GWs is first-order phase transitions (PTs)~\cite{Jedamzik:1999am,Liu:2021svg,Liu:2022lvz,Ellis:2019oqb,Baker:2021nyl,Cai:2024nln,Flores:2024lng}. 
	During first-order PTs, the effective potential exhibits at least two nondegenerate local minima. 
	In such a scenario, the metastable phase decays due to quantum tunneling or thermal fluctuations~\cite{Coleman:1977py,Callan:1977pt,Linde:1980tt}. 
	This process leads to the nucleation of numerous true vacuum bubbles, which expand and collide, converting vacuum energy into bubble walls and surrounding plasma~\cite{Caprini:2015zlo,Caprini:2019egz}. 
	These collisions generate a significant number of GWs, which have the potential to be detected by various GW observatories, such as LIGO/VIRGO/KAGRA~\cite{KAGRA:2021kbb}, Taiji~\cite{Ruan:2018tsw,Luo:2021qji}, TianQin~\cite{TianQin:2015yph}, Pulsar Timing Array~\cite{Xu:2023wog} and LISA~\cite{Barausse:2020rsu}.

	First-order PTs also play an important role in constructing the reheating process of the early Universe~\cite{Ai:2024cka,Barni:2024lkj,Buen-Abad:2023hex,Azatov:2024auq}. 
	Reheating is a crucial process that bridges the end of inflation and the onset of the hot big bang epoch. 
	During this stage, an interaction between the inflaton and the Standard Model particles is essential for the Universe to transition into the radiation-dominated era~\cite{Kofman:1997yn,Kofman:1994rk,Baumann:2009ds,Bassett:2005xm,Shtanov:1994ce}. 
	Specifically, the inflation field transfers its energy to ordinary particles, effectively heating the Universe and setting the initial condition for the subsequent cosmological evolution. 
	Extremely high temperatures during reheating create a favorable environment for high-energy physics theories, such as the grand unified theory (GUT) and string theory~\cite{Dvali:1995fb,Dvali:1998ct,Hu:2025xdt,Brdar:2019fur,Li:2020eun,Gubser:1998bc}. 
	These theories predict the existence of additional scalar fields with multiple vacuum states during inflation and reheating. 
	Quantum tunneling between these vacuum states could take place during reheating, influencing the dynamics of the early Universe. 
	Moreover, some studies suggest that during reheating, the coupling coefficient between the inflation field and the Standard Model is not constant but rather depends on certain scalar fields, potentially those associated with superstring theory or GUT~\cite{Dvali:1998ct}.
	
	In this paper, we focus on the assumption that the coupling coefficient between the inflaton and the Standard Model is not a constant but rather a function of a scalar field $\chi$. 
	We specifically explore the scenario in which $\chi$ undergoes  a first-order PT. 
	In general, the field value remains close to zero in the symmetric phase, while becomes nonzero in the broken phase. If the decay rate of inflaton is proportional to the $\chi$ field value, it will experience a sudden enhancement upon entering the true vacuum bubbles. Assuming the inflaton is a massive particle, the Universe would be matter-dominated prior to decay. The first-order PT of $\chi$ then triggers reheating, during which the inflaton gradually decays into radiation within the true vacuum of $\chi$. Moreover, the stochastic nature of bubble nucleation during this first-order PT leads to asynchronous decay rates across different spatial regions.
	During this stage, the energy densities of matter and radiation decay according to different laws. 
	The matter density decays as \(\rho_m \propto a^{-3}\)~(\(a\) is the scale factor of the Universe), while the radiation density decays more rapidly, following the law \(\rho_r \propto a^{-4}\). 
	As a result, in regions where the first-order PT occurs later, the energy density is higher at the end of reheating. 
	This induces energy density perturbations, which can induce GWs through the second-order coupling in the Einstein equations. 
	These GWs may be detectable by current and future GW detectors. Notably, due to the negligible vacuum energy of the PT, GWs are not produced through conventional mechanisms such as bubble collisions, sound waves, and turbulence~\cite{Kosowsky:1991ua,Kosowsky:1992vn,Huber:2008hg,Mazumdar:2018dfl,Caprini:2015zlo}.
	Instead, all GW signals originate from curvature perturbations. 
	This work also proposes a novel mechanism for the GW generation in the active research area of the matter-to-radiation transition in the early Universe~\cite{Bhaumik:2022pil,Papanikolaou:2020qtd,Aurrekoetxea:2023jwd,Balaji:2024hpu,Bhaumik:2020dor,Sui:2024grm}.
	For simplicity, we adopt natural units throughout this paper, setting \(c = 8\pi G = 1\).

	\section{curvature perturbations from PT during reheating }
	
	We postulate that the coupling coefficient between the inflation field and the particles in radiation is not constant but rather proportional to the field value of a scalar field $\chi$ with multiple vacuum states
	\begin{equation}    V_{\mathrm{int}}\sim\chi\phi\psi\overline{\psi}\,.
	\end{equation}
	At the end of inflation, the field $\chi$ resides in a false vacuum state $\chi_f=0$, which yields a vanishing interaction term between $\phi$ and $\psi$. 
	Instead of decay into radiation, the inflaton oscillates around the minimum of the quadratic potential, so the Universe is dominated by non-relativistic matter. 
	Since the false vacuum is unstable, the field $\chi$ will ultimately transition to the true vacuum state with $\chi_t> 0$ through a first-order PT, which enables efficient energy transfer from the inflation field to radiation.
	We assume that the decay rate of the inflaton decaying into radiation is \(\Gamma\), which depends on the field value of \(\chi\). 
	For simplicity, in this paper, we assume that \(\Gamma\) is a free parameter.
	
	In a first-order PT, the nucleation rate of true vacuum bubbles typically takes an exponential form~\cite{Enqvist:1991xw}
	\begin{equation}
		\gamma(t)=\gamma_0\exp(\beta t)\,.
	\end{equation}
	where $\gamma_0$ and $\beta$ are  parameters that depend on the specific models, which can be treated as constants. 
	The average probability $F(t)$ of the false vacuum is given by~\cite{Turner:1992tz}
	\begin{equation}
		\label{F}
		F(t)=\exp\left(-\frac{4\pi}{3}\int_{t_i}^{t}dt'\gamma(t')a^3(t')R^3(t,t')\right)\,,
	\end{equation}
	where $t_i$ marks the beginning of the PT for field $\chi$, and $R(t,t') := \int_{t'}^t a^{-1}(\overline{t}) d\overline{t}$ gives the comoving radius of true vacuum bubbles, with their expansion speed set to the light speed. Equation~\eqref{F} applies only when $t > t_i$; for $t < t_i$, the Universe remains entirely in the false vacuum state with $F = 1$.
	
	In this paper, the energy density \(\rho_{\chi}\) of the field \(\chi\) is much smaller than the energy density of the inflaton, \(\rho_{\phi}\), to ensure that inflation can occur normally and persist.
	Since the energy density of  $\chi$ is negligible, the Friedmann equation and the equations of motion at the end of inflation can be written as
	
	\begin{equation}
		\label{H}
		H^2=\frac{1}{3}(\rho_{r}+\rho_{m1}+\rho_{m2})\,,
	\end{equation}
	\begin{equation}
		\label{rhom1}
		\rho_{m1}(t)a^3=\rho_{i}F(t)\,,
	\end{equation}
	\begin{equation}
		\label{rhom2}
		\frac{d\left(\rho_{m2}a^3\right)}{dt}=-\frac{d\left(\rho_{m1}a^3\right)}{dt}-\Gamma\rho_{m2}a^3\,,
	\end{equation}
	\begin{equation}
		\label{rhor}
		\frac{d\left(\rho_{r}a^4\right)}{dt}=\Gamma\rho_{m2}a^4\,,
	\end{equation}
	where $H$ is the Hubble parameter and \(\rho_i\) is the energy density of the inflaton at time \(t = t_i\), \(t_i\) is defined as the moment after the end of inflation but before the start of the PT, corresponding to \(F(t_i) = 1\). $\rho_r$ represents the energy density of radiation, $\rho_{m1}$ and $\rho_{m2}$ represent the average energy densities of the inflaton within the false and true vacuum regions, respectively. Here, we neglect the energy densities corresponding $\chi$, including the bubble walls and the vacuum energy.
	In the false~(true) vacuum regions, the decay rate of the inflaton is vanishing~(nonzero), this condition results in the following consequence: 1. In Eq.~\eqref{rhom1}, the decrease of $\rho_{m1}a^{3}$ comes from the expansion of the bubbles , i.e., the decrease of $F(t)$ and the expansion of the Universe. 2. In Eq.~\eqref{rhom2}, the evolution of $\rho_{m2}a^{3}$ is determined by both the transition rate from $\rho_{m1}$ and the decay rate into radiation. 3. In Eq.~\eqref{rhor}, the increase of $\rho_r$ only depends on the decay of $\rho_{m2}$.
	\begin{figure*}
		\centering
		\begin{minipage}[t]{0.48\textwidth}
			\centering
			\includegraphics[width=\textwidth]{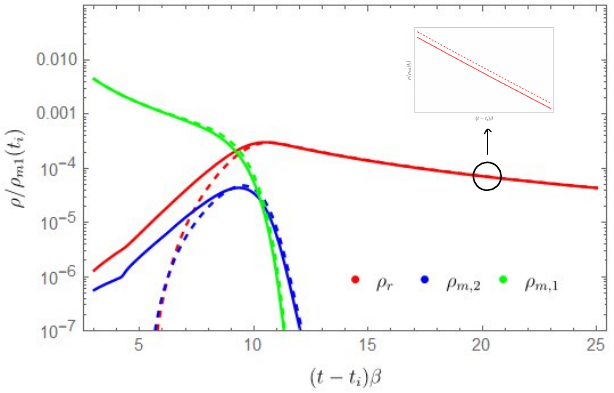}
			
			\label{fig:sub1}
		\end{minipage}
		\hfill
		\begin{minipage}[t]{0.48\textwidth}
			\centering
			\includegraphics[width=\textwidth]{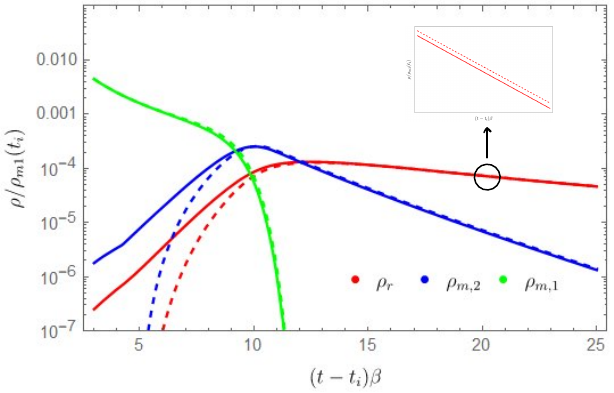}
			
			\label{fig:sub2}
		\end{minipage}
		\caption{This figure shows the evolution of the energy densities of radiation $\rho_r$, inflaton in false vacuum \(\rho_{m_1}\), and inflaton in true vacuum \(\rho_{m_2}\) over time when \(\beta/H_* = 100\) and \(\alpha = 4\) ($\alpha=1/4$) for left (right) panel. The solid lines describe the evolution of the background energy density, while the dashed lines describe the evolution of the energy density in the delayed decay regions.}
		\label{figevo}
	\end{figure*}
	Along with the expansion of the Universe, the energy densities of radiation and matter are respectively proportional to $\propto a^{-4}$ and $\propto a^{-3}$. 
	Consequently, regions with delayed true vacuum bubble nucleation of $\chi$ retain higher energy density at reheating completion compared to earlier-nucleating regions. 
	This asynchrony in vacuum decay across different Hubble horizons generates super-horizon curvature perturbations.
	To analyze these curvature perturbations, we employ a numerical method to calculate the density contrast $\frac{\delta \rho}{\rho}$. 
	We define the probability $P(t)$ that no true vacuum bubble has nucleated by a given time $t_n$ within a Hubble horizon as~\cite{Liu:2021svg,Zeng:2023jut}
	\begin{equation}
		P(t_n)=\exp\left(-\frac{4\pi}{3}\int_{t_i}^{t_n}\frac{a^3(t)}{a^3(t_e)}H^{-3}(t_e)\gamma(t)dt\right)\,,
	\end{equation}
	where $t_e$ is the time when the false vacuum in this Hubble horizon completely transitions to the true vacuum. 
	The total energy density within the Hubble horizon becomes larger after the PT for later $t_n$ with same $\Gamma$. 
	However, this delayed transition also corresponds to a lower probability of occurrence. 
	When considering a large region containing a sufficient number of Hubble horizons, we assume that the distribution of $\frac{\delta \rho}{\rho}$ follows a Gaussian distribution according to the central-limit theorem. 
	Therefore, we estimate the variance of the density perturbation using $t_v=t_n+\frac{1}{\Gamma}$, where $t_n$ satisfy $P(t_n) = 0.32$~\footnote{For perturbations with a Gaussian distribution, the probability of the regions exceeding the variance is approximately $0.16$~($\int_{\sigma}^{\infty} \frac{1}{\sigma\sqrt{2\pi}} e^{-\frac{x^2}{2\sigma^2}} dx \approx 0.16$). Since the delayed regions only covers half of the total space, we apply the requirement $P(t_n) = 0.32$.}.
	We then numerically calculate $\frac{\delta \rho}{\rho}$ by combining Eq.~\eqref{H}, \eqref{rhom1}, \eqref{rhom2}, and \eqref{rhor}. 
	
	\begin{figure}
		\centering
		\begin{minipage}[t]{0.98\linewidth}
			\includegraphics[width=\linewidth]{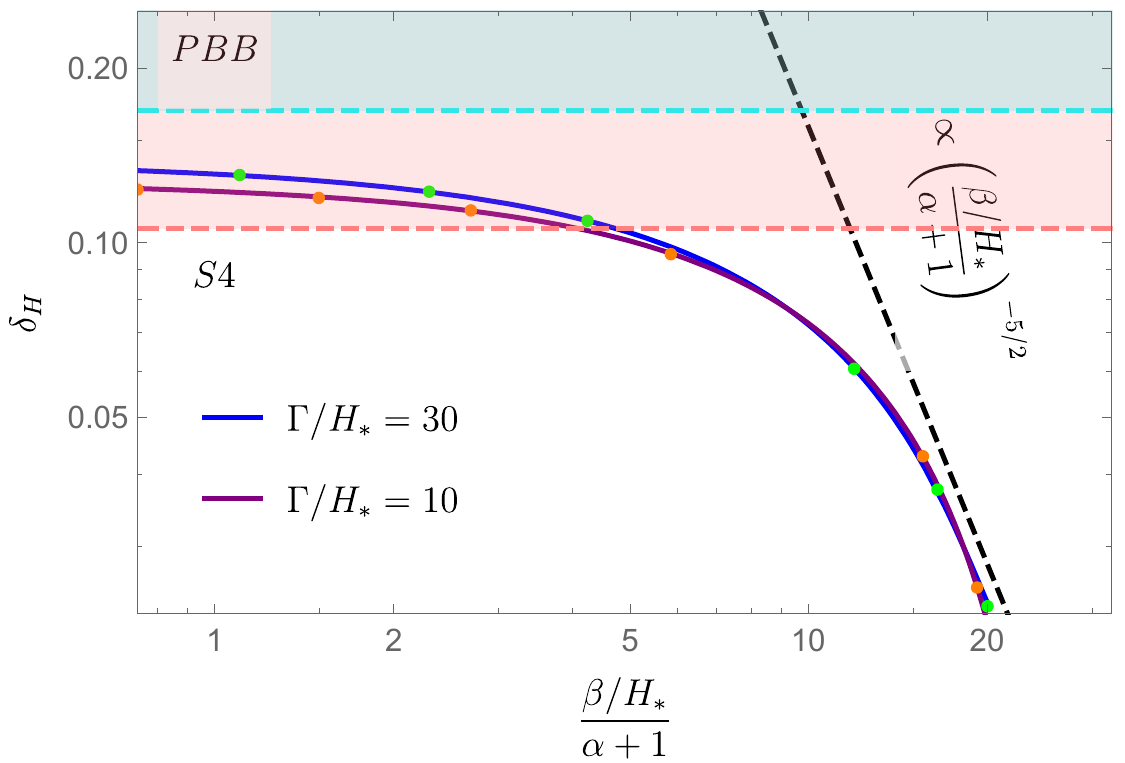}
		\end{minipage}
		\hfill
		\begin{minipage}[t]{0.98\linewidth}
			\includegraphics[width=\linewidth]{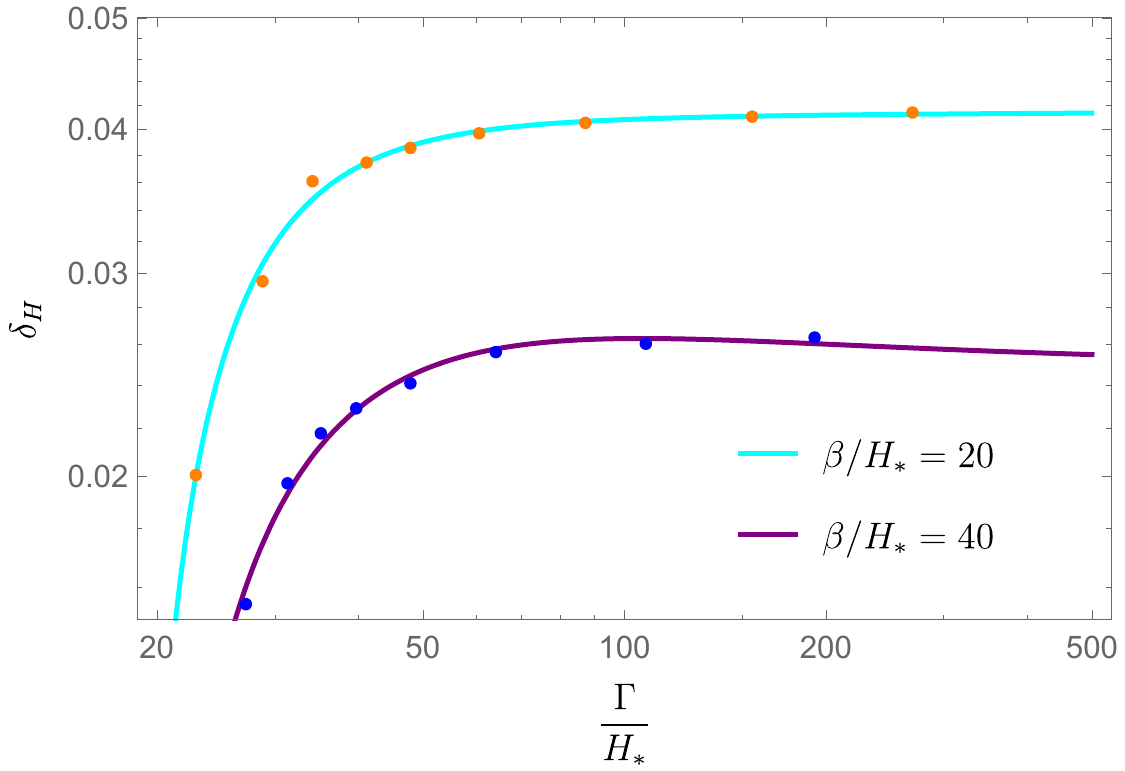}
		\end{minipage}
		\caption{This figure shows the numerical results of $\delta_H$ with differen parameters. 
			The upper (lower) panel describes the change of density perturbations with \(\frac{\beta/H_*}{\alpha+1}\) (\(\Gamma/H_*\)) when a fixed decay rate \(\Gamma\) (PT parameter \(\beta\)). 
			The orange and cyan dashed line in the upper panel represent the constraints of S4 and PBB on density perturbations, respectively.}
		\label{figdelta}
	\end{figure}

	Note that, in this work, we employ the delayed decay mechanism introduced in Refs.~\cite{Liu:2021svg,Liu:2022lvz}, which treats Hubble-scale regions as independently evolving domains and neglects the influence of bubble expansion from outside the horizon. 
	While more sophisticated approaches have recently been developed, notably the numerical methods in Refs.~\cite{Lewicki:2024ghw,Franciolini:2025ztf}, which elegantly incorporate the evolution and distribution of false vacuum fractions at horizon scales, with Ref.~\cite{Franciolini:2025ztf} additionally providing a rigorous treatment of metric perturbations—our scenario presents a fundamentally distinct physical picture. 
	Distinguished from previous studies that focus solely on the dynamics of first-order PTs themselves, we investigate a novel mechanism where the first-order PT modulates the inflaton decay rate. 
	In our work, the metric perturbations are very small, rendering their backreaction on field and fluid evolution negligible, which contrasts sharply with the large perturbations required for primordial black hole formation in earlier works~\cite{Jedamzik:1999am,Liu:2021svg,Baker:2021nyl,Baker:2021sno,Jung:2021mku,Hashino:2021qoq,Huang:2022him,He:2022amv,Kawana:2022olo,Gouttenoire:2023naa,Lewicki:2023ioy,Gouttenoire:2023bqy,Cai:2024nln,Flores:2024lng,Kanemura:2024pae,Ai:2024cka}. 
	Given these characteristics, the delayed decay approximation remains valid and provides a reliable estimation for our analysis in this new scenario.
	
	In general, the fundamental parameters are the PT rate and the PT strength. In our work, since the energy density of the scalar field is very small, only the PT rate plays a significant role. However, if the decay of the inflaton in the true vacuum is not instantaneous, the inflaton decay rate also emerges as a fundamental parameter. 
	The factors affecting large-scale density perturbations in this paper include not only the rate of the first-order PT but also the decay rate of the inflaton into radiation, \(\Gamma\). 
	Figure~\ref{figevo} shows the evolution of the energy densities of $\rho_r$, \(\rho_{m_1}\) and \(\rho_{m_2}\) where we have defined $\alpha:=\frac{\beta}{\Gamma}$.
	The reheating process is divided into two stages: first, due to the existence of the first-order PT, the inflaton transforms from the matter with small decay rate \(m_1\) to the matter with large decay rate \(m_2\); then, \(m_2\) decays into radiation. 
	We find that when \(\alpha\) is relatively large, significant radiation production only occurs after the effective generation of \(m_2\), and the Universe is briefly dominated by \(m_2\). 
	This is because a large \(\alpha\) indicates that the decay rate is much slower than the PT rate. 
	In this regime, the decay rate of \(m_2\) is insufficient to promptly convert the generated \(m_2\) into radiation, leading to the ``accumulation" of \(m_2\).  
	Conversely, when \(\alpha\) is small, the decay of the inflaton is very rapid. In this case, \(m_2\) produced by the PT decays into radiation quickly, preventing any substantial matter-dominated epoch.
	
	In Fig.~\ref{figdelta}, we present our numerical results of $\delta_H$ and observe that as \(\beta\) increases, the density perturbations gradually tend toward a power-law distribution. 
	To explain this result, we note that the characteristic time of this process can be approximately expressed as \(\beta^{-1} + \Gamma^{-1}\). We are concerned with density perturbations on large scales, so we divide the space into cubes with a spatial scale much larger than \(\beta^{-1} + \Gamma^{-1}\). 
	Different cubes can be considered to evolve independently, and within each cube, the density perturbation is proportional to \(\beta^{-1} + \Gamma^{-1}\). 
	Since the evolution between different cubes is independent, according to the Poisson distribution, the density perturbations on large scales satisfy \(\frac{\delta \rho}{\rho} \propto k^{3/2}\). The entire reheating process is a sub-horizon process; therefore, on the Hubble horizon scale, the variance after the end of reheating can be approximately expressed as \(\delta_H \propto \left( \frac{\beta/H_*}{\alpha + 1} \right)^{-5/2}\) and \(H_*\) denotes the Hubble parameter at the end of reheating.
	
	The decay rate \(\Gamma\) of the inflaton plays a crucial role here. 
	The lower panel of Fig.~\ref{figdelta} shows that as \(\Gamma\) gradually increases, the density perturbations increase rapidly, and the effects of \(\beta\) become increasingly pronounced. 
	This is because the generation of density perturbations relies on the fact that radiation dilutes faster than matter as the Universe expands. 
	The first-order PT can only determine the distribution of \(m_2\) and density perturbations emerge only after the inflaton decays into radiation. 
	As an illustrative explanation, when \(\Gamma\) is extremely small, it means that even if the Universe undergoes a first-order PT, the inflaton barely decays, and the Universe is completely dominated by matter, with density perturbations tending to zero. 
	On the other hand, when \(\Gamma \gg \beta\), it indicates that the decay of the inflaton in the true vacuum is highly efficient; after the true vacuum bubbles are generated, the matter inside will quickly decay into radiation, and the factor limiting the density perturbations at this time is mainly the rate of the PT.

	The power spectrum $\mathcal{P}_{\mathcal{R}}(k)$ of curvature perturbations is related to the Hubble-scale overdensity $\delta_H$ as follows~\cite{Carr:2009jm}
	\begin{equation}
		\label{deltaH}
		\delta_H^2=\frac{16}{81}\int_0^{\infty}\frac{dk}{k}\left(kR_H\right)^4
		W^2(k,R_H)\mathcal{P}_{\mathcal{R}}(k)\,,
	\end{equation}
	where $R_H$ is the Hubble radius at the end of the PT, and $W(k, R_H) = \exp\left(-k^2 R_H^2 \right)$ is a Gaussian window function. In the large-scale limit where $k \ll \beta^{-1}$, curvature perturbations follow the scaling $\mathcal{P}_{\mathcal{R}} \propto k^3$. Combining this with eq.~\eqref{deltaH}, we obtain the power spectrum of large-scale curvature perturbations
	\begin{equation}
		\label{PR}
		\mathcal{P}_{\mathcal{R}}(k)=34.5\delta_{H}^{2}\left(\beta/H_*,\alpha\right)R_H^3k^3\,,
	\end{equation}
	where the form of $\delta_H\left(\beta/H_*,\Gamma\right)$ is shown in Fig.~\ref{figdelta}. In order to ensure the reliability of the subsequent calculation results, we set a conservative cutoff scale $k_{\text{cut}}=R_H^{-1}$ for eq.~\eqref{PR}. 
	

	\section{GWs induced by scalars} \label{IGW}
	Curvature perturbations induced by the first-order PT begin to evolve after entering the horizon, ultimately giving rise to second-order GWs.
	Therefore, we expect this mechanism to generate a significant second-order GW signal, which could be observed by future GW detectors.
	Before exploring the specific characteristics of these second-order GWs, we first present the necessary formulas for their calculation~\cite{Kohri:2018awv}. 
	In the Newtonian gauge, the perturbed metric of a Friedmann-Robertson-Walker Universe is given by~\cite{Mukhanov:2005sc}
	\begin{equation}
		\label{ds}
		ds^2=a^2(\eta)\left(-\left(1 + 2\Phi\right)d\eta^2+\left(\delta_{ij}-2\Phi\delta_{ij}+h_{ij}\right)dx^idx^j\right)\,,
	\end{equation}
	where $\eta$ denotes the conformal time, and we neglect vector perturbations, first-order GWs, and anisotropic stress.
	
	At the second-order expansion of the Einstein equation, the scalar and tensor perturbations of the metric are mutually coupled. The equation of motion for tensor perturbations in the Fourier space can be presented as
	\begin{equation}
		h_{\textbf{k}}''+2\mathcal{H}h_{\textbf{k}}'+k^2h_{\textbf{k}} = 4S_{\textbf{k}}\,,
	\end{equation}
	where $S_{\textbf{k}}$ is the source term originating from first-order scalar perturbations.

	Once GWs generated, the interaction between GWs and the background plasma is negligible, and the GW energy density scales with the cosmic expansion as $a^{-4}$. Leveraging entropy conservation, the present day GW energy spectrum, $\Omega_{\text{GW},0}$, can be expressed in terms of GW energy spectrum $\Omega_{\text{GW}}$ in the strongly radiation dominated (sRD) era as~\cite{Espinosa_2018,Inomata:2020tkl}
	\begin{equation}
		\Omega_{\text{GW},0}h^2(k)=0.39\left(\frac{g_0}{106.75}\right)^{-1/3}\Omega_{\text{r},0}h^2\Omega_{\text{GW}}(\eta_{\text{eq}},k)\,,
	\end{equation}
	where $\Omega_{\text{r},0}h^2\sim4.18\times10^{-5}$ represents the energy density fraction of radiation evaluated at present~\cite{Planck:2018vyg}. 
	In this paper, $g$ is used to denote the effective relativistic degrees of freedom and $\eta_{\text{eq}}$ represents the time when the densities of matter and radiation are equal in the standard big - bang cosmological model.
	
	We are concerned with the evolution of scalar perturbations during the radiation - dominated era of the Universe. 
	Using the Green function method, the GW energy spectrum $\Omega_{\text{GW}}$ of induced GWs can be evaluated by following the integral~\cite{Kohri:2018awv}
	\begin{widetext}
		\begin{equation}
			\begin{split}
				\Omega_{\text{GW}}(\eta,k)=&99.2\delta_H^4(\beta/H_*,\alpha)\int_{0}^{\infty}dv\int_{|1-v|}^{1+v}du\left(\frac{4v^2-(1+v^2-u^2)^2}{4uv}\right)^2\left(\frac{ku}{k_{\text{cut}}}\right)^3\left(\frac{kv}{k_{\text{cut}}}\right)^3\left(\frac{3}{4u^3v^3}\right)^3\left(u^2+v^2-3\right)^2\\
				&\left(\pi^2\left(u^2+v^2-3\right)^2\Theta\left(u+v-\sqrt{3}\right)+\left(-4uv+\left(u^2+v^2-3\right)\ln\left|\frac{3-(u+v)^2}{3-(u-v)^2}\right|\right)^2\right)\,.
			\end{split}
			\label{gw}
		\end{equation}
	\end{widetext}
	
	\begin{figure*}
		\centering
		\begin{minipage}[t]{0.48\textwidth}
			\centering
			\includegraphics[width=\textwidth]{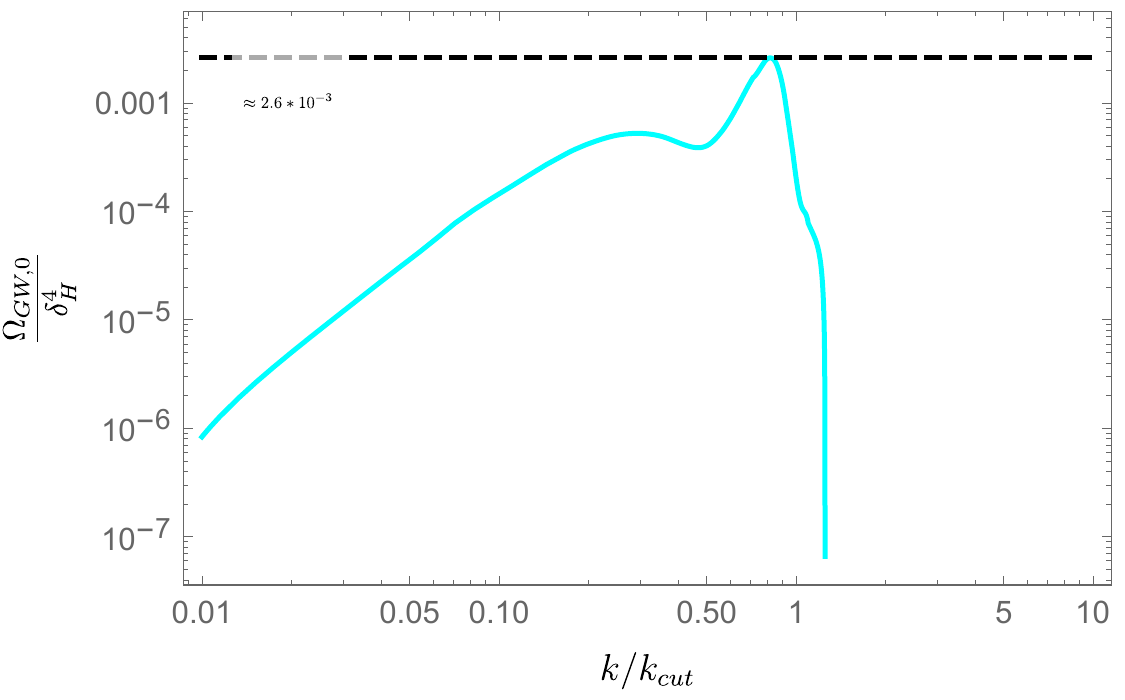}
			
			\label{fig:sub1}
		\end{minipage}
		\hfill
		\begin{minipage}[t]{0.48\textwidth}
			\centering
			\includegraphics[width=\textwidth]{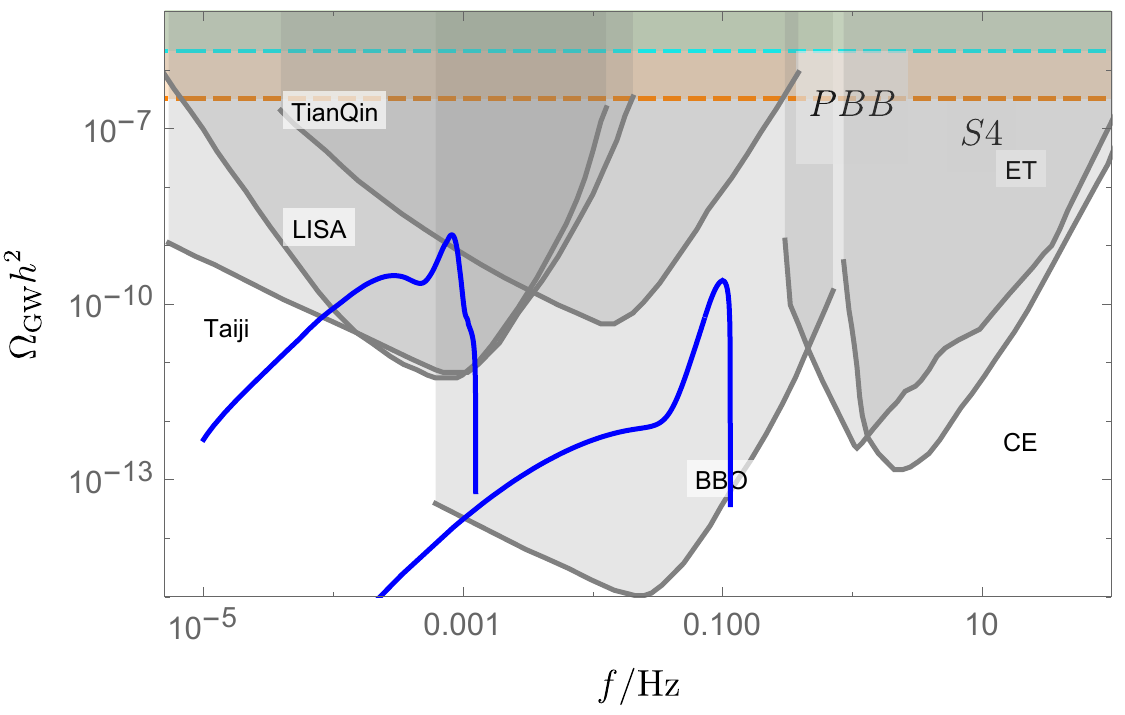}
			
			\label{fig:sub2}
		\end{minipage}
		\caption{This figure we plot the energy spectrum of second-order scalar-induced GWs. 
			The left panel shows the component of the GW energy spectrum independent of density perturbations $\delta_H$, denoted as \(\Omega_{\text{GW,0}}/\delta^4_H\), where the black dashed line represents the peak of this component. 
			The right panel presents the GW energy spectrum under the parameter choices of \(\beta/H_* = 47\), \(\alpha = 4/3\), \(k_{\text{max}}/k_{\text{cut}} = 100\), and \(k_{\text{cut}} = 0.001\ \text{Hz}\), where the left curve describes the GWs induced by density perturbations $\delta_H$, and the right curve describes the second-order GWs amplified by the rapid transition from the early matter-dominated era to the radiation-dominated era. 
			The cyan and orange dashed lines represent the constraints of PBB and S4 on GWs, respectively.}
		\label{figgw}
	\end{figure*}
	
	The left panel of Fig.~\ref{figgw} shows the results of $\Omega_{\mathrm{GW}}$ by solving Eq.~\eqref{gw}. First, we find that the peak frequency of the GWs lies near \(k_{\text{cut}}\) and the peak value is determined by $\delta_H$ which roughly satisfy~$\Omega_{\text{GW,peak}}\approx2.6\times10^{-3}\delta^4_H$.
	A distinctive feature of our scenario is that $\Omega_{\mathrm{GW}}$ is not suppressed by the PT strength parameter, defined as the ratio of the released vacuum potential energy to the radiation energy density, allowing the scalar-induced GWs to be considerably stronger.

	From the cosmological point of view, GWs constitute of dark radiation and can be parameterised by an correction of the effective degree of freedom, $\Delta N_{\text{eff}}=N_{\text{eff}}-N^{\text{SM}}_{\text{eff}}$.
	A higher $N_{\text{eff}}$ can delay radiation-to-matter equality and change the size of the sound horizon, which can leave features on CMB anisotropies, baryon acoustic oscillations and Big-Bang nucleosynthesis. In this way, we can get the GW upper bounds inferred from cosmological constraints on $N_{\text{eff}}$.
	The limit given by the current and future observations on $\Delta N_{\text{eff}}$ is~\cite{Cang:2022jyc}
	\begin{equation}
		\Delta N_{\text{eff}}=\begin{cases}
			0.175, &Planck+\text{BAO}+\text{BBN}\,\text{(PBB)}\,,\\
			0.027, &\text{CMB Stage IV}\,\text{(S4)}\,.
		\end{cases}
	\end{equation}
	This gives the upper bound on GWs density at $95\%$ C.L.
	\begin{equation}
		\label{gwmax}
		\Omega_{\text{GW,bound}}=\begin{cases}
			2.11\times10^{-6}, &\text{PBB}\,,\\
			3.25\times10^{-7}, &\text{S4}\,.
		\end{cases}
	\end{equation}
	Through these observational constraints on the dark radiation energy density
    \begin{equation}
		\label{gwmax}
		\delta_{\text{H}}\lesssim\begin{cases}
			0.174, &\text{PBB}\,,\\
			0.106, &\text{S4}\,,
		\end{cases}
	\end{equation}
    and we can rule out scenarios with smaller values of \(\frac{\beta/H_*}{\alpha+1}\). 
    As illustrated in Fig.~\ref{figdelta}, the generated density perturbations remain consistent with the PBB and S4 bounds only when \(\frac{\beta/H_*}{\alpha+1}\) is sufficiently large.

	The Universe is matter-dominated before the inflaton decays, and the rapid transition from the early matter-dominated era to the radiation-dominated era causes the amplification of second-order GWs on small scales, which is a sub-horizon process~\cite{Assadullahi:2009nf}.
	During the matter-dominated phase, density perturbations on subhorizon scales experience continuous growth. 
	The rapid transition into radiation-dominated era enhances the scalar-induced GW signal at subhorizon scales~\cite{Assadullahi:2009nf,Sui:2024grm,Bhaumik:2020dor}. 
	Consequently, our model predicts two GW peaks located on either side of \(k = k_{\text{cut}}\): one at large scales (\(k < k_{\text{cut}}\)) arising from the super-horizon curvature perturbations characterized by \(\delta_H\), and another at small scales (\(k > k_{\text{cut}}\)) resulting from the gravitational amplification during the matter-dominated epoch. 
	Notably, the large-scale GW peak is not constrained by the duration of the matter-dominated era or the requirement for small-scale structure formation, making it more accessible to detection.

	The right panel of Fig.~\ref{figgw} presents our main results for the parameter choices \(\beta/H_* = 47\), \(\alpha = 4/3\), and \(k_{\text{cut}} = 0.01\ \text{Hz}\). 
	The left GW peak arises from second-order GWs sourced by the large-scale curvature perturbations \(\delta_H\). 
	Since \(k_{\text{cut}}\) is determined by the Hubble parameter and scale factor at the end of reheating, observing this peak enables us to constrain the energy scale of the reheating epoch.
	The right GW peak originates from density perturbations that grew within the horizon during the matter-dominated era, where the sharp peak is a consequence of the rapid transition to radiation domination~\cite{Pearce:2023kxp,Inomata:2019zqy,Inomata:2020lmk}. 
	We set \(k_{\text{max}}/k_{\text{cut}} = 100\), where this ratio characterizes the duration of the early matter-dominated era.
    Note that the value of $k_{\text{max}}$ is model-dependent, which is mainly determined by the duration of the inflaton-dominated era. Here we choose \(k_{\text{max}}/k_{\text{cut}} = 100\) as an example to illustrate the numerical results of $\Omega_{\mathrm{GW}}$ from the rapid transition to radiation domination, which is showed in Fig.~\ref{figgw}.
    Typically, this peak follows the relation \( \Omega_{\text{GW}} \propto \left(k_{\text{max}}/k_{\text{cut}}\right)^7 \) in the infrared slope~\cite{Inomata:2019ivs}.
    Small-scale density perturbations within the cosmic horizon undergo gravitational growth as the Universe evolves. 
    An excessively prolonged early matter-dominated era would invalidate the framework of perturbation theory. Consequently, a common constraint in relevant studies is \( k_{\text{max}}/k_{\text{cut}} < 450 \)~\cite{Assadullahi:2009nf}.
    
	Remarkably, both GW peaks can potentially be observed simultaneously by next-generation GW observatories.
	Such a distinctive double-peak spectral signature serves as a smoking-gun feature of this mechanism, enabling clear discrimination from other GW sources in the early Universe.

	\section{Conclusions} \label{conclusion}
	
	In this paper, we investigate the scenario where the reheating process is mudulated by a scalar field that undergoes a first-order PT, where the coupling coefficient between the inflaton and radiation is not a constant, but rather modulated by the aforementioned scalar field. 
	During reheating, the stochastic nature of the PT causes variations in reheating times across different Hubble horizons, thereby inducing large-scale curvature perturbations. 
	To explore this process, we have applied a semi-analytical method to determine the generation of density perturbations.  
	Our results demonstrate that both the PT parameter \(\beta\) and the inflaton decay rate \(\Gamma\) play crucial roles in shaping the density perturbations. Specifically, when for large value of \(\frac{\beta/H_*}{\alpha+1}\), density perturbations exhibit a power-law distribution, which is consistent with theoretical expectations. 

	We emphasize a key distinction from other first-order PT studies: since the energy density of the field undergoing the PT is negligible, the conventional PT strength parameter does not appear in our framework. Instead, the inflaton decay rate \(\Gamma\) directly controls the efficiency of density perturbation generation. As \(\Gamma\) approaches zero, the PT-induced density perturbations vanish rapidly; conversely, when \(\Gamma/\beta \gg 1\), the density perturbations are dominated primarily by the PT dynamics.  
	Furthermore, since the inflaton dominates the cosmic energy budget prior to reheating, this mechanism naturally generates substantial density perturbations and correspondingly strong GW signals.
	
	After the completion of the PT, the Universe enters the radiation-dominated era. The curvature perturbations induced by first order PT start to evolve upon reentering the horizon, leading to the generation of second-order GWs. 
	Our calculation results show that the peak value of the induced GW signal is determined by the PT parameter \(\frac{\beta}{H_*}\) and \(\alpha\), and has a quartic relationship with \(\delta_H\). 
    The Hubble parameter at the end of reheating determines the peak frequency of the GWs.  
	In addition, the rapid transition from the early matter-dominated era to the radiation-dominated era amplifies the second-order GWs on small scales which gradually increases in the matter-dominated era. 
	Such a double-peak GW signal can serve as a characteristic signature of this mechanism, distinguishing it from other GW sources. 
	
	Although the context discussed in this paper is the reheating process, our results are applicable to other studies involving the early matter-dominated era, such as those on Q-balls and primordial black holes. Compared with reheating, the energy scales corresponding to these scenarios are lower, so the density perturbations are subject to stricter constraints from the CMB~\cite{Jinno:2021ury,Chluba:2015bqa,Lucca:2019rxf} or UCMH abundance~\cite{Clark:2015sha,Clark:2015tha}.  

	The reheating process is often accompanied by the production of dark matter, and variations in reheating times can lead to perturbations in the large-scale distribution of dark matter~\cite{Hall:2009bx}. 

	\begin{acknowledgments}
		This work is supported in part by the National Key Research and Development Program of China Grants No. 2020YFC2201501 and No. 2021YFC2203002, in part by the National Natural Science Foundation of China Grants No. 12105060, No. 12147103, No. 12235019, No. 12075297 and No. 12147103, in part by the Science Research Grants from the China Manned Space Project with NO. CMS-CSST-2021-B01, in part by the Fundamental Research Funds for the Central Universities.
	\end{acknowledgments}

	\bibliography{citeLib}
\end{document}